\documentclass[aps, prd,twocolumn,showpacs,superscriptaddress,nofootinbib]{revtex4-1}

\usepackage{braket}
\usepackage{amsfonts}
\usepackage{amsmath}
\usepackage{amssymb}
\usepackage{amsthm}
\usepackage{bm}
\usepackage{dcolumn}
\usepackage{epsfig}
\usepackage{graphicx}
\usepackage{graphics}
\usepackage[utf8]{inputenc}
\usepackage{latexsym}
\usepackage{rotating}
\usepackage[colorlinks=true]{hyperref}
\usepackage[usenames]{color}
\usepackage{yfonts}
\usepackage{float}
\usepackage{ucs}
\usepackage{xspace} 
\usepackage{mathrsfs}
\usepackage{booktabs}
\usepackage[toc,page]{appendix}
\usepackage{array}
\usepackage[normalem]{ulem}
\usepackage{tikz}

\newcommand{\be}{\begin{equation}}
\newcommand{\ee}{\end{equation}}

\begin{document}
	\title{Prospects for Fundamental Physics with LISA}

        \author{Enrico Barausse}
	\affiliation{SISSA, Via Bonomea 265, 34136 Trieste, Italy and INFN
Sezione di Trieste, and IFPU - Institute for Fundamental Physics of the Universe, Via Beirut 2, 34014 Trieste, Italy}
\affiliation{
CNRS, UMR 7095, Institut d'Astrophysique de Paris, 98 bis Bd Arago, 75014 Paris, France.}
	
        \author{Emanuele Berti}
        \affiliation{Department of Physics and Astronomy, Johns Hopkins University, 3400 N. Charles Street, Baltimore, Maryland 21218, US}

        \author{Thomas Hertog}
	\affiliation{Institute for Theoretical Physics, KU Leuven, Celestijnenlaan 200D, B-3001 Leuven, Belgium.}

        \author{Scott A.~Hughes}
	\affiliation{Department of Physics and Kavli Institute for Astrophysics and Space Research, Massachusetts Institute of Technology, Cambridge, Massachusetts 02139, USA.}
		
	\author{Philippe Jetzer}
	\affiliation{Physik-Institut, Universit\"at Z\"urich, Winterthurerstrasse 190, 8057 Zu\"rich, Switzerland.}

	\author{Paolo Pani}
	\affiliation{Dipartimento di Fisica, ``Sapienza'' Universit\`a di Roma \& Sezione INFN Roma1, Piazzale Aldo Moro 5, 00185, Roma, Italy.}
	
	\author{Thomas P.~Sotiriou}
	\affiliation{School of Mathematical Sciences \& School of Physics and Astronomy, University of Nottingham, University Park, Nottingham, NG7 2RD, United Kingdom.}

	\author{Nicola Tamanini}
	\affiliation{Max-Planck-Institut f\"ur Gravitationsphysik, Albert-Einstein-Institut, Am M\"uhlenberg 1, 14476 Potsdam-Golm, Germany.}

	\author{Helvi Witek}
        \affiliation{Department of Physics, King's College London, The Strand, WC2R 2LS, London, UK and Department of Physics and
        University of Illinois at Urbana-Champaign, Urbana, Illinois 61801, USA.}

	\author{Kent Yagi}
	\affiliation{Department of Physics, University of Virginia, Charlottesville, Virginia 22904, USA.}
	
	\author{Nicol\'as Yunes}
	\affiliation{eXtreme Gravity Institute, Department of Physics, Montana State University, Bozeman, MT 59717, United States and
        University of Illinois at Urbana-Champaign, Urbana, Illinois 61801, USA.} 

      \maketitle
        
{T. Abdelsalhin$^{1}$, 
A. Achucarro$^{2}$, 
K. V. Aelst$^{3}$, 
N. Afshordi$^{4}$, 
S. Akcay$^{5}$, 
L. Annulli$^{69}$, 
K. G. Arun$^{7}$, 
I. Ayuso$^{132}$, 
V. Baibhav$^{9}$, 
T. Baker$^{10}$
% ...
\\
{\footnotesize$^{1}$Dipartimento di Fisica, ``Sapienza'' Universit\`a di Roma \& Sezione INFN Roma1, Piazzale Aldo Moro 5, 00185, Roma, Italy.}\\
{\footnotesize$^{2}$LISA NL Group, University of Leiden, Netherland}\\
{\footnotesize$^{3}$Laboratoire Univers et Theories, France}\\
{\footnotesize$^{4}$Perimeter, Canada}\\
{\footnotesize$^{5}$University of Jena Relativity Group, Germany}\\
{\footnotesize$^{6}$LUTH, Observatoire de Paris, France}\\
{\footnotesize$^{7}$Chennai Mathematical Institute, India}\\
{\footnotesize$^{8}$eXtreme Gravity Institute, Montana State University,Illinois Relativity Group, USA}\\
{\footnotesize$^{9}$ Department of Physics and Astronomy, Johns Hopkins University, 3400 N. Charles Street, Baltimore, Maryland 21218, US}\\
{\footnotesize$^{10}$Queen Mary University of London, UK}\\
{\footnotesize$^{11}$Institut d'Astrophysique de Paris, France}\\
{\footnotesize$^{12}$Instituto de Astrof\'isica e Ci\^encias do Espaco, Lisboa, Portugal}\\
{\footnotesize$^{13}$ICC Durham, UK}\\
{\footnotesize$^{14}$Dipartimento di Fisica e Astronomia Galileo Galilei, Universit\`a di Padova, 35131 Padova Italy and
INFN, Sezione di Padova, 35131 Padova Italy}\\
{\footnotesize$^{15}$LISA NL Group, UvA, Netherland}\\
{\footnotesize$^{16}$D\'epartement de Physique Th\'eorique and Center for Astroparticle Physics,
Universit\'e de Gen\`eve, 24 quai Ansermet, CH--1211 Gen\`eve 4, Switzerland}\\
{\footnotesize$^{17}$Astrophysics, University of Oxford, Oxford OX1 3RH, UK}\\
{\footnotesize$^{18}$Institut de Ci\`encies del Cosmos, Universitat de Barcelona, Mart\'i i Franqu\`es 1, 08028 Barcelona, Spain }\\
{\footnotesize$^{19}$Ariel Early Universe Physics}\\
{\footnotesize$^{20}$Institut de physique th\'eorique, Universit\'e Paris
Saclay, CEA, CNRS, 91191 Gif-sur-Yvette, France}\\
{\footnotesize$^{21}$Laboratoire APC, Paris, France}\\
{\footnotesize$^{22}$LISA NL Group, RU physics, Netherland}\\
{\footnotesize$^{23}$IUCAA}\\
{\footnotesize$^{24}$IRFU Saclay, France}\\
{\footnotesize$^{25}$Portsmouth University, UK}\\
{\footnotesize$^{26}$Department of Mathematics and Statistics, University of Otago, New Zealand}\\
{\footnotesize$^{27}$CEICO - Czech Academy of Sciences, Prague, Czechia}\\
{\footnotesize$^{28}$Van Swinderen Institute for Particle Physics and Gravity, University of Groningen, the Netherlands}\\
{\footnotesize$^{29}$SYRTE, Observatoire de Paris, France}\\
{\footnotesize$^{30}$Theoretical Particle Physics and Cosmology Group, Physics Department, King's
College London, University of London, Strand, London WC2R 2LS, UK}\\
{\footnotesize$^{31}$University College London, UK}\\
{\footnotesize$^{32}$LISA NL Group, University of Twente, Netherland}\\
{\footnotesize$^{33}$Theoretical Physics Group and Quantum Alberta,
Department of Physics and Astronomy, University of Lethbridge,
4401 University Drive, Lethbridge, Alberta T1K 3M4,
Canada}\\
{\footnotesize$^{34}$Departments of Astronomy and Physics, University of Arizona, Tucson, Arizona 85721, USA}\\
{\footnotesize$^{35}$Johannes Gutenberg University Mainz, Institute for Physics, Germany}\\
{\footnotesize$^{36}$LISA NL Group, Nikhef, Netherland}\\
{\footnotesize$^{37}$University of Edinburgh, UK}\\
{\footnotesize$^{38}$GGC Gravity Group}\\
{\footnotesize$^{39}$Instituto de Estructura de la Materia - CSIC, Madrid, Spain}\\
{\footnotesize$^{40}$Institute for Theoretical Physics, University of
Amsterdam, PO Box 94485, 1090GL Amsterdam}\\
{\footnotesize$^{41}$Universita L'Aquila, Italy}\\
{\footnotesize$^{42}$Department of Physics, University of Illinois at
Urbana-Champaign, Urbana, IL 61820, USA}\\
{\footnotesize$^{43}$Jet Propulsion Laboratory, Caltech, USA}\\
{\footnotesize$^{44}$Cornell, USA}\\
{\footnotesize$^{45}$Kepler Center T\"ubingen (KeCeT), Germany}\\
{\footnotesize$^{46}$Universit\'e Libre de Bruxelles, CP 231, B-1050 Brussels, Belgium}\\
{\footnotesize$^{47}$Uppsala University, Sweden}\\
{\footnotesize$^{48}$Imperial College London, UK}\\
{\footnotesize$^{49}$Instituto Superior Tecnico, Portugal}\\
{\footnotesize$^{50}$Case Western Reserve University, USA}\\
{\footnotesize$^{51}$University of Sheffield, UK}\\
{\footnotesize$^{52}$NZ Astrostatistics and GR, University of Auckland, New-Zealand}\\
{\footnotesize$^{53}$Cardiff University Gravity Exploration Institute, UK}\\
{\footnotesize$^{54}$Wilbur Wright College, Chicago, USA}\\
{\footnotesize$^{55}$Institute of Cosmology and Gravitation, University of Portsmouth, UK}\\
{\footnotesize$^{56}$MSLU, Russia}\\
{\footnotesize$^{57}$Japanese working group for LISA science, YITP, Kyoto University}\\
{\footnotesize$^{58}$Taiwan Consortium for Gravitational Wave Research, National Tsing Hua University, Taiwan}\\
{\footnotesize$^{59}$Szeged Gravity Group, University of Szeged, Hungary}\\
{\footnotesize$^{60}$School of Physics and Astronomy and Institute for Gravitational Wave Astronomy, University of Birmingham, Birmingham, B15 2TT, UK}\\
{\footnotesize$^{61}$Haggard Research Group at McGill University}\\
{\footnotesize$^{62}$Taiwan Consortium for Gravitational Wave Research, Academia Sinica, Taiwan}\\
{\footnotesize$^{63}$Shanghai Astronomical Observatory, CAS, China}\\
{\footnotesize$^{64}$Aix Marseille Univ, CNRS/IN2P3, CPPM, Marseille, France}\\
{\footnotesize$^{65}$Glasgow LISA Science, UK}\\
{\footnotesize$^{66}$University of Stavanger, Norway}\\
{\footnotesize$^{67}$Humboldt State University Gravitational Research Laboratory}\\
{\footnotesize$^{68}$LIGO Laboratory, Massachusetts Institute of Technology, Cambridge, Massachusetts 02139, USA}\\
{\footnotesize$^{69}$CENTRA, Departamento de Fisica,
Instituto Superior T\'ecnico - IST, Universidade de Lisboa - UL,
Avenida Rovisco Pais 1, 1049-001 Lisboa, Portugal}\\
{\footnotesize$^{70}$University of California, Davis, USA}\\
{\footnotesize$^{71}$Cornell University, USA}\\
{\footnotesize$^{72}$Lohrmann Observatory, Technische Universit??t Dresden, Germany}\\
{\footnotesize$^{73}$University of Wisconsin - Madison, USA}\\
{\footnotesize$^{74}$Massachusetts Institute of Technology, USA}\\
{\footnotesize$^{75}$Taiwan Consortium for Gravitational Wave Research, National Taiwan Normal University, Taiwan}\\
{\footnotesize$^{76}$Taiwan Consortium for Gravitational Wave Research, Tamkang University, Taiwan}\\
{\footnotesize$^{77}$CSU Fullerton Gravitational-Wave Physics and Astronomy Center, USA}\\
{\footnotesize$^{78}$University Granada, Spain}\\
{\footnotesize$^{79}$eXtreme Gravity Institute, Montana State University, USA}\\
{\footnotesize$^{80}$Complutense University of Madrid \& IPARCOS, Spain}\\
{\footnotesize$^{81}$Japanese working group for LISA science, Ryukoku University}\\
{\footnotesize$^{82}$Technical University of Liberec, Czech Republic}\\
{\footnotesize$^{83}$Department of Physics, Aristotle University of Thessaloniki, Thessaloniki, Greece}\\
{\footnotesize$^{84}$CAUP and IA-Porto, Portugal}\\
{\footnotesize$^{85}$Harvard-Smithsonian Center for Astrophysics, USA}\\
{\footnotesize$^{86}$LISA NL Group, UU physics, Netherland}\\
{\footnotesize$^{87}$Harvard Physics, USA}\\
{\footnotesize$^{88}$NCSA, University of Illinois at Urbana-Champaign Department of Physics, USA}\\
{\footnotesize$^{89}$ISS-Sci, Romania}\\
{\footnotesize$^{90}$Japanese working group for LISA science, Kyushu University}\\
{\footnotesize$^{91}$Kavli Institute for Astronomy and Astrophysics, Peking University, Beijing 100871, China}\\
{\footnotesize$^{92}$Georgia Tech, USA}\\
{\footnotesize$^{93}$Institute of Space Sciences (ICE, CSIC), Campus UAB,
Carrer de Can Magrans s/n, 08193 Cerdanyola del Vall\`es
(Barcelona), and Institute of Space Studies of Catalonia (IEEC),
Carrer del Gran Capit\`a, 2-4, Edifici Nexus, despatx 201,
08034 Barcelona, Spain}\\
{\footnotesize$^{95}$Universit\'e d'Orl\'eans CNRS (LPC2E), France}\\
{\footnotesize$^{96}$Japanese working group for LISA science, ICRR, University Tokyo}\\
{\footnotesize$^{97}$Japan instrument group, Japan}\\
{\footnotesize$^{98}$Japanese working group for LISA science, Kyoto University}\\
{\footnotesize$^{99}$APC and Institut d’Astrophysique de Paris, CNRS and Sorbonne Universit\'es, UMR
7095, 98 bis bd Arago, 75014 Paris, France}\\
{\footnotesize$^{100}$University of Minnesota / CEICO, USA}\\
{\footnotesize$^{101}$Tata Institute of Fundamental Research, Homi Bhabha Road,
 Mumbai 400005, India}\\
{\footnotesize$^{102}$Kyoto University, Japan}\\
{\footnotesize$^{103}$UFL: LISA group at University of Florida, USA}\\
{\footnotesize$^{104}$NZ Astrostatistics and GR, University of Canterbury, New-Zealand}\\
{\footnotesize$^{106}$University of Texas at Austin, USA}\\
{\footnotesize$^{107}$UC Berkeley, USA \& IPhT Saclay, France}\\
{\footnotesize$^{108}$Instituto de Fisica Teorica UAM/CSIC, Universidad Autonoma
  de Madrid, Cantoblanco 28049 Madrid, Spain}\\
{\footnotesize$^{109}$Laboratoire de Physique et Chimie de l'Environnement et de l'Espace LPC2E UMR7328, Universit\'e d'Orl\'eans, CNRS, F-45071 Orl\'eans, France}}\\
{\footnotesize$^{110}$Center for Particle Cosmology,
Department of Physics and Astronomy, University of Pennsylvania,
209 S. 33rd St.,
Philadelphia, PA 19104, USA }\\
{\footnotesize$^{111}$TRIUMF, Canada}\\
{\footnotesize$^{112}$Van Swinderen Institute, University of Groningen, 9747 AG,
Groningen, Netherlands}\\
{\footnotesize$^{113}$University of Chinese Academy of Sciences (UCAS)}\\
{\footnotesize$^{114}$Kavli Institute for Cosmological Physics, The University of Chicago, USA}\\
{\footnotesize$^{115}$ESOC - European Space Operations Centre,
D-64293 Darmstadt, Germany}\\
{\footnotesize$^{116}$Max Planck Institute for Gravitational Physics, Potsdam, Germany}\\
{\footnotesize$^{117}$SISSA, International School for Advanced Studies, Via Bonomea 265, 34136 Trieste, Italy}\\
{\footnotesize$^{118}$Laboratoire Astroparticule et Cosmologie, CNRS,
 Universit\'e Paris Diderot, 75013, France}\\
{\footnotesize$^{119}$School of Physics and Astronomy and Institute for Gravitational Wave Astronomy, University of Birmingham, Birmingham, B15 2TT, UK}\\
{\footnotesize$^{120}$CIDMA and Aveiro University, Portugal}\\
{\footnotesize$^{121}$Victoria University of Wellington, New Zealand}\\
{\footnotesize$^{122}$Physics Department, Technion, 3200003 Haifa, Israel}\\
{\footnotesize$^{123}$Center for Computational Astrophysics, Flatiron Institute, 162 5th Ave, New York, NY 10010}\\
{\footnotesize$^{124}$Department of Physics, University of
California, Santa Barbara, CA 93106}\\
{\footnotesize$^{125}$Institute for Theoretical Physics, KU Leuven, Celestijnenlaan 200D, B-3001 Leuven, Belgium}\\
{\footnotesize$^{126}$School of Physics and Astronomy, University of Nottingham, Nottingham, NG7 2RD, UK}\\
{\footnotesize$^{127}$Theoretical Physics Department, CERN, 1 Esplanade des Particules, CH-1211 Geneva 23, Switzerland}\\
{\footnotesize$^{128}$INFN, Sezione di Padova, via Marzolo 8, I-35131, Padova, Italy}\\
{\footnotesize$^{129}$Department of Applied Mathematics, The University of Western Ontario, N6A 5B7, London, Ontario, Canada}\\
{\footnotesize$^{130}$Swansea University, UK}\\
{\footnotesize$^{131}$Theoretical Physics Group, Department of Physics,
Kuwait University, P.O. Box 5969, Safat 13060, Kuwait}\\
{\footnotesize$^{132}$IA, Portugal}\\
\date{\today}


In this paper, which is of programmatic rather than
quantitative nature, we aim to further delineate and sharpen
the future potential of the LISA mission in the area of
fundamental physics. Given the very broad range of topics
that might be relevant to LISA, we present here a sample of
what we view as particularly promising fundamental physics
directions. We organize these directions through a
``science-first'' approach that allows us to classify how
LISA data can inform theoretical physics in a variety of
areas. For each of these theoretical physics classes, we
identify the sources that are currently expected to provide
the principal contribution to our knowledge, and the areas
that need further development. The classification presented
here should not be thought of as cast in stone, but rather
as a fluid framework that is amenable to change with the
flow of new insights in theoretical physics.  



\section{Introduction}

Several of the deepest open questions in fundamental physics involve gravity in one way or another. These include the classical and quantum dynamics of black holes, a detailed understanding of the expansion and structure formation history in cosmology, and of course the fundamental nature of gravity and spacetime itself. 

Gravitational wave (GW) observations have an enormous potential to inform and to falsify theoretical work in these areas, leading to exciting prospects for a fruitful interplay between fundamental theory and observation. On the one hand GWs give us access to largely unexplored regions of the universe that are dark, such as the immediate environment of black holes and the earliest phases of large-scale structure formation, and to regions where light cannot penetrate, such as the very early universe. On the other hand GWs provide a source of information that complements conventional astronomy and cosmology, enabling a ``multi-messenger'' approach, thereby paving the way for a deeper understanding. 

The observation of long-wavelength GWs with LISA \cite{Lisa} is particularly promising as a probe of fundamental physics. Potential examples are anomalies in the data related to gravitational parity violation, which could provide a hint toward a resolution of the baryogenesis problem. Other anomalies related to violations of the Equivalence Principle or Lorentz invariance could produce modifications in the dispersion relation of matter or horizon-scale modifications in black hole physics due to quantum gravity effects. Observations of the dispersion relation of GWs could constrain a large class of modified theories, which include massive gravity models that attempt to explain the late-time acceleration of the universe, as well as other Lorentz-violating theories (such as Einstein-\ae ther or Horava gravity), whose renormalizability makes them attractive candidates for quantum gravity.

The observational input that LISA will provide will also be complementary to that following from ground-based GW observations \cite{Ligo1,Ligo2,Ligo3}, carried out by LIGO, VIRGO and KAGRA, because the target sources are qualitatively different. LISA will observe GWs at much lower frequencies than ground-based instruments, allowing for the measurement of an entirely different class of sources: supermassive black hole mergers, EMRIs, galactic binaries, and stochastic GW backgrounds. Some of these sources, such as supermassive black hole mergers, will lead to extremely loud signals, with signal-to-noise ratios in the thousands, that will allow for a deep search of anomalies. Other classes of sources will lead to signals that may not be very loud, such as the EMRIs, but that will nonetheless be extremely complex with lots of amplitude and phase modulations, allowing for the search of qualitatively different anomalies. Moreover, weak signals may allow for tests of General Relativity (GR) that are statistically enhanced by the large number of events, and which might therefore be competitive against single events with extremely large signal-to-noise ratios. LISA observations also complement future GW observations via
pulsar timing arrays and the B-mode polarization in the cosmic microwave background, both of which probe an even lower GW frequency range.


The goal of this paper is to identify and scientifically motivate a sample of topics in fundamental physics beyond the current standard models of particle physics, gravity and cosmology that we view are particularly relevant for the LISA scientific community. These topics are of interest to several Working Groups (WG) organized within the LISA Consortium: the Fundamental Physics, the Cosmology, the Astrophysics, and the Waveform-Modeling WGs, each of which approaches these from a  different, complementary angle.
We stress that here we shall discuss all the topics in a qualitative way, as precisely the more quantitative
aspects will be subject to detailed investigations and are thus as such not yet know. 
Once new results will be available the relevance of certain topics will of course change and new
ones, not yet known, might arise. Thus this paper has to be seen as a first step with the aim
to somehow coordinate the effort needed towards formulating a realistic assessment in the area
of the fundamental physics feasible with LISA. Thus this paper will definitively rise more questions
than giving answers.  

\begin{figure}[ht]
\vspace{0.3cm}
\begin{center}
\begin{tabular}{cc}
  \includegraphics[clip=true,width=9.5cm,clip=true]{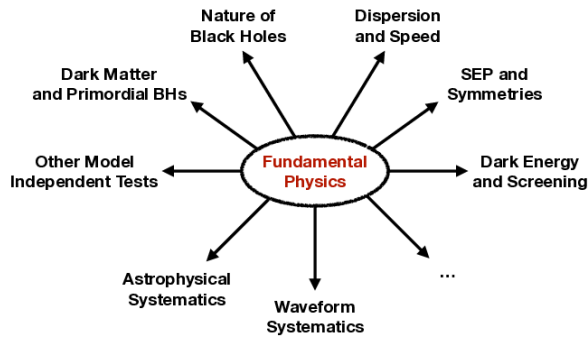}
\end{tabular}
\end{center}
\vspace{-0.6cm}
\caption{\label{fig0}~A taxonomy of LISA-related topics in fundamental physics. Each of them is discussed in a separate section in this paper. The ellipses stand for topics that may be considered in the future.}
\end{figure}

This initiative should be viewed not as an exhaustive classification but rather as a warmup for a more comprehensive and detailed account in the future. Our discussion will be organized in a \emph{science-first} approach. That is, instead of first thinking about sources of GWs, we will first think about the theoretical physics that could be learned with LISA, irrespective of the source class. Of course, any such list will be, by definition, incomplete, and perhaps more importantly, only a snapshot of the interests of the field at the time of writing. One should thus think of the classes we will identify below as fluid, subject to change in the future, as the winds of physics start blowing in a different direction. With this caveat in mind, we identify the science drivers presented in Fig.~\ref{fig0}, with each science driver defined and discussed in much more detail in each of the sections that follows. This classification implicitly assumes that work must be done in three main areas: theoretical development, waveform generation, and data analysis, with different drivers currently at different levels of development. 

With the classes declared, we will then sub-organize each class with sub-classes, following a source classification approach. For the purpose of this document we will identify 
six different source sub-classes: 
\begin{itemize}
\setlength\parskip{0.0cm}
\setlength\itemsep{0.2cm}
\item Supermassive Black Hole Binaries (SMBHBs): Coalescences with mass ratio larger than $10^{-1}$ and total masses in $(10^{5},10^{7}) M_{\odot}$.
\item Intermediate-Mass Black Hole Binaries (IMBHBs): Coalescences with mass ratio larger than $10^{-1}$ and total masses in $(10^{2},10^{5}) M_{\odot}$.
\item Extreme mass-ratio and intermediate mass-ratio inspirals (EMRIs and IMRIs): Coalescences with mass ratios in $(10^{-6},10^{-3})$ and $(10^{-3},10^{-1})$, and total masses in $(10^{3},10^{7}) M_{\odot}$.
\item Stellar origin BH binaries (SOBHBs): Inspirals with sufficiently low total mass (e.g.~in $(50,500) M_{\odot}$) such that they could be detected both by LISA and second- or third-generation ground-based detectors.
\item Galactic Binaries: White dwarf or neutron star binary inspirals within the Milky Way that produce nearly monochromatic signals. 
\item Stochastic Backgrounds: Cosmological sources of GWs that produce a stochastic background. 
\end{itemize}
Of these source sub-classes, SMBHBs with accretion disks, SOBHBs in nuclear galactic disks~\cite{Bartos:2016dgn}, and galactic binaries are expected to produce strong and coincident electromagnetic signals. By no means ought this to be thought of as final, since LISA could always detect sources that nobody expected. Inversely, we make no statements in this document about the astrophysical rates of these events, or even whether all of these will be detectable with LISA, as this will depend on the noise of the actual detector (see e.g.~\cite{Moore:2019pke}).

The rest of this paper is organized as follows. 
Section~\ref{sec:mod-disp} discusses modified dispersion relations and the speed of gravity. 
Section~\ref{sec:violSEPandSym} describes violations of the Equivalence Principle and violations of other fundamental symmetries.
Section~\ref{sec:natureBH} covers tests of the nature of black holes.
Section~\ref{sec:DE-screening} discusses dark energy and screening. 
Section~\ref{sec:DM-PBH} describes dark matter and primordial black holes. 
Section~\ref{sec:other-model-indep-tests} summarizes ideas for other model-independent tests. 
Section~\ref{sec:astro-sys-and-stacking} discusses astrophysical systematics, while Sec.~\ref{sec:waveform-sys} covers waveform systematics.
Section~\ref{sec:conclusions} summarizes and concludes with an outlook to the future. 
Henceforth, we employ geometric units when needed, in which $G = 1 = c$ and we follow the conventions of~\cite{Misner:1973cw}. 

\section{Modified Dispersion Relations and the Speed of Gravity}
\label{sec:mod-disp}

According to Einstein's theory, GWs obey the dispersion relation $\omega^{2} = k_{i} k^{i}$, with the contraction done with the flat Euclidean metric. This then immediately implies that the group and the phase velocity of GWs are the speed of light. Modified theories of gravity, in particular those that attempt to unify quantum mechanics and GR, sometimes lead to different dispersion relations of the form
\be
\label{eq:mod-disp}
\omega^{2} = k_{i} k^{i} + \frac{m_{g}^{2}}{\hbar^{2}} + A (k_{i} k^{i})^{\alpha}\,,
\ee
where $m_{g}$ is a hypothetical mass for the graviton, $\alpha \in \mathbb{R} \backslash \{0\}$ determines the type of modification introduced, and $A$ controls its magnitude. This expression should be thought of as approximate, in the limit that  $m_{g}^{2}/\hbar^{2} \ll k^{2}$ and $A \ll k^{2-\alpha}$.

The parameterization of the correction to the propagation of GWs presented in Eq.~\eqref{eq:mod-disp} is obviously not unique, and other parameterizations have been considered in the literature, especially in the context of cosmology~\cite{Ezquiaga:2017ekz,Baker:2017hug,Ezquiaga:2018btd,Sakstein:2017xjx}.

A commonly used parameterization is 
\be
\label{eq:mod-disp-new}
\omega^{2} + i H \omega \left(3 + \alpha_{M}\right) = \left(1 + \alpha_{T}\right) k_{i} k^{i}\,,
\ee
where we are here considering waves propagating in a cosmological background with Hubble parameter $H$. A more detailed discussion on these assumptions and consequences for black-hole properties can be found in~\cite{Tattersall:2018map}.

Clearly, $\alpha_{T} = A$ when $\alpha = 1$, and it controls the speed of GWs. The parameter $\alpha_{M}$ is not included in Eq.~\eqref{eq:mod-disp}, and it controls the rate of dissipation of GWs (see e.g.~\cite{Alexander:2017jmt}). Both parameterizations have advantages and disadvantages. For example, Eq.~\eqref{eq:mod-disp} allows one to constrain a kinematical graviton mass, while Eq.~\eqref{eq:mod-disp-new} does not, whereas Eq.~\eqref{eq:mod-disp-new} allows one to test the rate of GW dissipation, while Eq.~\eqref{eq:mod-disp} does not. 

A modification of this type clearly leaves an imprint on the GWs that arrive on Earth, but this imprint is due to modifications in the propagation of the waves, and not modifications in their generation. One can think of this modification as a correction to the graviton propagator in quantum field theory language. Given this, one can in principle modify any wave generation scheme by simply modifying the way the GWs propagate from the source to the detector on Earth in vacuum. For a more detailed review of the way this modification affects the response function, see~\cite{Mirshekari:2011yq,Yunes2013}. 

The best systems to constrain these modifications are those that are as far away as possible from Earth, which reduces to SMBHBs (see e.g.~\cite{Chamberlain:2017fjl}). This is because modifications to the propagation of GWs accumulate with distance traveled. In addition, constraints on the mass of the graviton are also enhanced for supermassive systems because the correction scales with the chirp mass. For $\alpha > 1$, however, the opposite is true, with constraints deteriorating as an inverse power of the chirp mass~\cite{Mirshekari:2011yq,Yunes2013}. 

A confirmation of the dispersion relation of GR could place constraints on theories with extra dimensions or quantum-inspired Lorentz violation (both predict $\alpha=2$), and on modified gravity models that attempt to explain the late-time acceleration of the universe (that predict $\alpha=0$ and $m_{g} \neq 0$). An important distinction should be made, however, regarding the speed of gravity. Typically, when we refer to the speed of gravity, we mean the constant coefficient in front of the first term on the right-hand side of Eq.~\eqref{eq:mod-disp}. This constant cannot be measured precisely with only GWs detected with a space-based instrument or multiple ground-based detectors~\cite{Blas:2016qmn,Cornish:2017jml}. Instead, its precise determination requires an electromagnetic coincident observation~\cite{Monitor:2017mdv}. Therefore, constraints on the speed of gravity are typically only possible with neutron star binaries or black hole-neutron star binaries that induce an electromagnetic signal when considering ground-based detectors, and with supermassive black hole mergers when considering space-based detectors~\cite{DalCanton:2019wsr}. 
EMRIs in which a neutron star falls into a supermassive black hole will not lead to tidal disruption outside the horizon of a supermassive black hole due to the latter's mass, thereby decreasing the chances to generate detectable electromagnetic signals. LIGO-Virgo observations have already constrained the speed of gravity to better than one part in $10^{15}$~\cite{Monitor:2017mdv}. 

The current status on the development of models for this type of test is mostly complete. Not included above is the possibility to have an anisotropic dispersion relation, for example due to preferred frame effects; a classification of such anisotropic effects from the viewpoint of effective field theory can be found in~\cite{Kostelecky:2016kfm,  Mewes:2019dhj}. Modifications to the propagation of the GW can be implemented \emph{a posteriori} after the wave generation problem has been solved. Typically, the propagation modification is modular and can be implemented on any model in typically a straightforward way. Additional work could be devoted to verifying the validity of such an implementation in regimes where the stationary phase approximation breaks down. 

\section{Violations of the Equivalence Principle and Fundamental Symmetries}
\label{sec:violSEPandSym}

The equivalence principle has been a guiding principle in gravitation for centuries. In Newton's original formulation it refers to the equivalence between inertial mass and gravitational mass. The definition has evolved to avoid reference to the mass in a relativistic setup and it has also been expanded to include other principles. We refer the reader to \cite{Will:1993ns} for a thorough discussion. 

The {\em Weak Equivalence Principle} (WEP), or simply universality of free fall, postulates that if an {\em uncharged test body} is placed at an initial event in spacetime and given an initial velocity there, then its subsequent trajectory will be independent of its structure and composition. The {\em Einstein Equivalence Principle} postulates that the WEP, Local Lorentz Invariance (LLI), and Local Position Invariance (LPI) hold true. LLI requires that the outcome of any local non-gravitational test experiment is independent of the velocity of the freely-falling apparatus. LPI requires that the outcome of said experiment is independent of where and when it is performed. The {\em Strong Equivalence Principle} (SEP) extends the WEP to self-gravitating bodies and the LLI and LPI to {\em any} experiment. In the context of the WEP,  LLI and LPI are intimately linked with Lorentz symmetry and universality of couplings in the standard model, as there is reference to non-gravitational experiments. In the SEP, LLI and LPI can be seen as manifestations of Local Lorentz symmetry and universality of couplings in the standard model and gravitation. 

Einstein's theory is the only known gravity theory that, minimally coupled to the
standard model (SM) of particle physics, satisfies the SEP.  So a null test of the SEP is usually considered as a confirmation of GR. Though there is no definitive proof that GR is unique in this respect, it is relatively straightforward to argue that generic deviations of GR and/or the standard model would indeed lead to SEP violations. The first step is to realize that deviation from the standard model or GR generically implies the existence of new fundamental fields. This is rather obvious for the SM, while for GR it is implied by Lovelock's theorem \cite{Lovelock:1972vz,Sotiriou:2015lxa}.\footnote{Theories with auxiliary fields manage to circumvent Lovelock theorem without introducing new dynamical degrees of freedom, but they have serious shortcomings \cite{Flanagan:2003rb,Barausse:2007pn,Pani:2012qd,Pani:2013qfa}.} To circumvent the latter and put together a gravity theory other than GR, one needs to either explicitly introduce new fields that couple non-minimally to gravity, allow for higher-order field equations, allow for more than 4 spacetime dimensions, or give up diffeomorphism invariance (general covariance). While from a fundamental physics perspective these options might be distinct, the latter three options lead back to having additional fields if one adopts a more phenomenological viewpoint. Indeed, higher-order equations mean more degrees of freedom generically, higher-dimensional models can be compactified down to 4 dimensions plus additional fields, and diffeomorphism invariance can be restored by rewriting the same theory using a larger field content (Stueckelberg trick). See \cite{Sotiriou:2015lxa} for a discussion.

It should now be clear why deviations from GR will generically lead to violations of the SEP. Extra fields imply additional interactions. If black holes and compact stars carry nontrivial configurations of the new fields then they can experience these new interactions. This would lead to a direct violation of the SEP. Additionally, the new interaction can also lead to additional GW polarizations (e.g.~longitudinal dipolar emission) that affect orbital dynamics. Moreover, in the context of Lorentz-violating gravity theories, these new fields could select a preferred direction in spacetime, thereby leading to violations of the LLI. Finally, the configurations of these extra fields could depend on the spacetime location, thereby leading to violations of LPI.

Based on the above, and the fact that LISA's primary binary sources contain at least one black hole, LISA's ability to constrain violations of the SEP is intimately related to its ability to test the structure of black holes  in a given beyond-GR scenario. This will be covered in some detail in the next section, so we will postpone the technical discussion until then.
A theoretical limitation comes from no-hair theorems. Black holes in theories that are covered by no-hair theorems will 
not carry additional charges\footnote{The term charge is used here loosely, to refer to a non-trivial configuration of 
some new field, as has become customary  in the relevant literature.} and hence they will not exhibit any additional 
interactions. In such cases there will not be violation of the SEP, though one might hope to detect deviations from GR 
through quasinormal ringing (see Sec.~\ref{sec:natureBH}), since the latter can be affected by nontrivial couplings even 
when the background solution is a GR black hole~\cite{Barausse:2008xv,Molina:2010fb,Tattersall:2018nve}. 

No-hair theorems \cite{Hawking:1972qk,Sotiriou:2011dz,Hui:2012qt,Silva:2017uqg,Sotiriou:2015pka,Herdeiro:2015waa} and ways to circumvent them have been heavily scrutinized in (generalized) scalar-tensor theories. Hairy black holes are known to exist in theories where the scalar couples to higher-order curvature invariants, such as the Gauss--Bonnet invariant~\cite{Campbell:1991kz,Kanti:1995vq,Yunes:2011we,Sotiriou:2013qea,Sotiriou:2014pfa,Antoniou:2017acq,Benkel:2016kcq,Benkel:2016rlz}
or the Pontryagin density \cite{Yunes:2009hc,Stein:2014xba}. Such couplings can arise from 
low-energy effective actions for quantum gravity candidates~\cite{Metsaev:1987zx,Maeda:2009uy,Jackiw:2003pm,Green:1987mn,Ashtekar:1988sw}. A coupling with the Pontryagin density breaks parity invariance as well and, hence, studying theories that include it offers a way to test parity violation in gravity\footnote{Technically speaking, the non-minimal coupling in the action is through the product of a pseudoscalar and the Pontryagin density and the action is parity invariant.  The solutions exhibit parity violations.} \cite{Jackiw:2003pm,Alexander:2009tp}. It has recently been shown that a suitable coupling with the Gauss--Bonnet invariant can lead to spontaneous scalarization of black holes \cite{Doneva:2017bvd,Silva:2017uqg}, {\em i.e.}~black holes that have hair only if they lie in a critical mass range. 

Modeling binaries numerically in theories with higher order curvature invariants is a challenging task that has just started~\cite{Okounkova:2017yby,Witek:2018dmd,Okounkova:2019dfo}. 
Constraints based on orbital effects of dipolar scalar emission seem to suggest that LISA is not likely to be competitive relative to  future ground based detectors, when it comes to binaries with similar masses \cite{Witek:2018dmd}. This is because the scalar charge is determined by the curvature, i.e., the inverse of the BH mass. Thus, somewhat counter-intuitively, the scalar charge
due to higher curvature corrections is small for supermassive BHs. At the same time large mass ratios seem to induce a strong scalar flux during the merger~\cite{Witek:2018dmd}, thus making IMRIs/EMRIs potentially interesting sources.

Nonetheless, the fact that LISA is probing a different mass range is quite important because, in a theory agnostic perspective, it will probe a different regime. Moreover, the effects of dipolar emission are not the only avenue for obtaining constraints. For example, it is not yet clear if modeling EMRIs within these theories could lead to stronger bounds. 

Instead of focusing on theories for which no-hair theorems do not apply, one can try to circumvent no-hair theorems by violating their assumptions. A non exhaustive list of examples that can be astrophysically interesting include the following: 
long-lived scalar clouds powered by superradiance provided that the scalar field has a very small mass~\cite{Arvanitaki:2010sy,Brito:2015oca};
truly hairy solutions branching off at the onset of the instability~\cite{Herdeiro:2014goa}
and long-lived scalar ``wigs'' even around non-rotating BHs~\cite{Barranco:2012qs,Barranco:2013rua};
time-dependent scalar configuration \cite{Babichev:2013cya,Clough:2019jpm,Hui:2019aqm,Berti:2013gfa}, if they can be supported by some non-trivial cosmological boundary conditions; 
hair induced by a non-trivial initial scalar field configuration~\cite{Healy:2011ef}; 
hair induced by matter in the vicinity of a black hole \cite{Cardoso:2013fwa,Hui:2019aqm}, though one might have to reach very high levels of precision; black holes in Einstein-dilaton-Maxwell theories~\cite{Hirschmann:2017psw,Julie:2018lfp,Astefanesei:2019pfq} and in generalized Proca theories~\cite{Minamitsuji:2017aan,Babichev:2017rti,Heisenberg:2017xda,Kase:2018owh,Rahman:2018fgy}.

We close this section with some additional remarks on testing LLI. Observations of gamma-ray bursts (see e.g.~\cite{AmelinoCamelia:1997gz}) and the binary neutron star merger GW170817 have already provided a strong constraint on Lorentz symmetry in gravity, in terms of a double-sided bound on the speed of GWs to a part in $10^{15}$ \cite{Monitor:2017mdv}. However, Lorentz-violating theories have multidimensional parameter spaces and generically exhibit additional polarizations \cite{Sotiriou:2017obf}. The speed of these new polarizations remains virtually unconstrained~\cite{Gumrukcuoglu:2017ijh,Oost:2018tcv}. 

More generally, Lorentz symmetry is essential to the very definition, and hence the structure, of a black hole. Lorentz-violating theories, such as Einstein-aether theory \cite{Jacobson:2000xp,Jacobson:2008aj} and Ho\v rava gravity \cite{Horava:2009uw,Blas:2009qj,Sotiriou:2010wn}, will typically exhibit superluminal or even instantaneous propagation of signals \cite{Blas:2011ni,Bhattacharyya:2015uxt}. This implies that black holes in these theories will have a different causal structure, featuring multiple horizons, corresponding to modes that travel at different speed \cite{Eling:2006ec,Barausse:2011pu}, or a new type of {\em universal} horizon that blocks all modes irrespective of propagation speed \cite{Barausse:2011pu,Blas:2011ni,Bhattacharyya:2015gwa}. This different horizon structure is expected to leave an imprint in the quasi-normal spectrum of these black holes. Moreover, it is important to emphasize that black holes in Lorentz-violating theories will in general have hair (although see~\cite{Ramos:2018oku}): for the field that can be thought of as breaking Lorentz symmetry can never be trivial.  Hence, though advances in modelling would have to be made to obtain quantitative estimate for constraints one could get with LISA, it should be clear that there is a strong potential for constraining Lorentz violations. It is worth emphasizing that bounds coming from testing the dispersion relation of GWs discussed in the previous section, are complementary to those discussed here.

\section{Tests of the Nature of Black Holes}
\label{sec:natureBH}

General Relativistic black holes are the simplest objects in the Universe. Within GR, 
the mass, angular momentum, and electric charge (although the latter is thought to be negligible\footnote{The well-motivated assumption that black holes should be uncharged could also be tested; see~\cite{Bozzola:2019aaw} for work in this direction.}) of an 
astrophysical black holes uniquely define its entire multipolar structure and its quasinormal-mode~(QNM) spectrum. This implies 
that the infinite number of multipole moments or of the QNM of a Kerr black hole are related to each 
other, a property that allows to perform multiple null-hypothesis tests of the no-hair 
theorem in various complementary ways.

In any extension of GR, the QNMs of a black hole can be parametrized as
\begin{eqnarray}
  \omega &=&\omega^{\rm Kerr}+\delta \omega \,,\\
  \tau   &=&\tau^{\rm Kerr}  +\delta \tau \,,
\end{eqnarray}
where $\omega$ and $\tau$ are the frequency and damping time of the mode.
Owing to its very large signal-to-noise ratio (SNR) in the ringdown phase of supermassive black holes, LISA will be able to perform 
\emph{black hole spectroscopy} by measuring several QNMs for a single event up to redshift 
$z=10$, thus constraining beyond-GR deviations, encoded in $\delta \omega$ and $\delta\tau$, with 
a precision that cannot be reached by any (present and future) ground-based GW 
interferometer~\cite{Dreyer:2003bv,Berti:2005ys,Berti:2016lat}.
Such modified ringdown frequencies for black holes arise, for example, in quadratic gravity.
The dilation or axion coupled to, respectively, the Gauss--Bonnet invariant or the Pontryagin density lead to truly hairy black holes.
Consequently, their ringdown exhibits a modulation due to the superposition of gravitational-led and scalar-led 
modes~\cite{Molina:2010fb,Witek:2018dmd,Okounkova:2019dfo}.

Another way to test the no-hair theorem is by measuring the multipole moments of a black hole. 
Several gravity theories beyond GR predict deformations of the Kerr metric, which 
result in a different multipolar structure,
\begin{eqnarray}
  {M}_\ell &=& {M}_\ell^{\rm Kerr} +\delta {M}_\ell \,, \\
  {S}_\ell &=& {S}_\ell^{\rm Kerr} +\delta {S}_\ell \,,
\end{eqnarray}
where $M_\ell$ and $S_\ell$ are the mass and current multipole moments, whereas $\delta 
{M}_\ell$ and $\delta {S}_\ell$ are theory-dependent corrections to the Kerr moments, which 
might even include cases in which the geometry breaks the equatorial symmetry~\cite{Raposo:2018xkf}.
The most accurate way to measure the multipole moments is by probing accurately the 
spacetime near a compact object. LISA will be able to do so by detecting the coalescence of comparable-mass 
binaries at high SNR~\cite{Kastha:2018bcr} and, especially, by detecting EMRIs that can perform millions 
of orbits around the central supermassive objects before 
plunging~\cite{Barack:2006pq,AmaroSeoane:2007aw,Babak:2017tow}. This allows 
to put constraints as stringent as $\delta M_2/M^3<10^{-4}$ on the quadrupole moment of 
the central supermassive object~\cite{Ryan:1995wh,Ryan:1997hg,Sotiriou:2004bm,Barack:2006pq,Babak:2017tow}.
Furthermore, the EMRI dynamics is also very sensitive to the existence of extra degrees 
of freedom predicted in almost any modified theory of gravity. In the presence of extra 
polarizations or dipolar radiation, the inspiral would proceed faster than in GR, 
affecting the GW phase in a detectable way~\cite{Cardoso:2011xi,Yunes:2011aa,Barausse:2016eii}. 
Furthermore, it is also possible that low-frequency modes (which are absent in GR) 
can be excited during the inspiral, leaving a characteristic imprint in the waveform.

While in GR the geodesic motion of 
a test particle around a Kerr black hole is fully integrable, in some theories the orbital motion 
might even be chaotic~\cite{Cardenas-Avendano:2018ocb}, which would also impact on the structure of the 
QNMs~\cite{Pappas:2018opz}. The presence of chaos in EMRIs is expected to be encoded in chaotic plateaus 
in the temporal evolution of the fundamental frequencies of the motion~\cite{Apostolatos:2009vu,LukesGerakopoulos:2010rc,Lukes-Gerakopoulos:2014tga,Contopoulos:2011dz,Lukes-Gerakopoulos:2017jub}. Exactly how such plateaus affect
the Fourier transform of the GW response function is not yet clear. Given that EMRIs are expected
to lead to low signal-to-noise ratio sources, new data analysis techniques may be required to test for the presence 
of chaotic motion in EMRI GWs. 

Finally, black holes and neutron stars might be just two ``species'' of a larger family of 
compact objects. More exotic species are theoretically predicted in extensions of GR, in 
the presence of beyond-standard model fields minimally coupled to gravity (e.g., boson 
stars~\cite{Jetzer:1991jr,Liebling:2012fv}), in Grand Unified Theories in the early Universe (e.g., cosmic strings), 
and in exotic states of matter. Several arguments~\cite{Cardoso:2019rvt} predict 
horizonless compact objects (e.g., fuzzballs, gravastars, and dark stars), or new physics 
at the horizon scale (e.g., firewalls). 

GW observations provide a unique discovery 
opportunity in this context, since exotic matter or dark matter might not interact 
electromagnetically, and any electromagnetic signal from the surface of a compact object 
might be highly redshifted. Example GW signatures include finite-size and 
tidal-deformability effects, as well as critical behavior, in the inspiral which might be detectable for 
highly-spinning SBHBs~\cite{Krishnendu:2017shb,Maselli:2017cmm,Addazi:2018uhd,Compere:2017hsi}
and for EMRIs~\cite{Pani:2019cyc}, a different QNM 
spectrum~\cite{Chirenti:2007mk,Pani:2009ss,Glampedakis:2017cgd}, excitation of internal oscillation modes of the 
object, and the 
presence of a surface instead of an event horizon. A smoking gun of the latter would be 
the presence of ``GW echoes'' in the post-merger GW signal of a 
coalescence~\cite{Cardoso:2016rao,Cardoso:2016oxy,Cardoso:2017cqb,Bueno:2017hyj} 
which are absent in the classical black hole picture in which the horizon is a one-way membrane. 
Supermassive black hole coalescence detectable by LISA with SNR as high as a few thousands will allow to 
constrain models of supermassive exotic compact objects in almost the entire region of 
their parameter space~\cite{Testa:2018bzd}.

Theoretical studies of the viability of such horizonless objects are advancing in recent years~\cite{Compere:2019ssx,Giddings:2019ujs}. 
One key criterion
for their viability is their stability~\cite{Eperon:2016cdd}. 
Horizonless objects will be unstable to light-ring (non-linear) instabilities~\cite{Keir:2014oka,Cardoso:2014sna,Cunha:2017qtt}
and to an ergoregion instability~\cite{Maggio:2018ivz}. The former can in principle be avoided if one modifies GR
or if the horizonless object is made of matter that violates the energy conditions.  The latter can be avoided if
one allows the surface of the horizonless object to be partially absorbing~\cite{Maggio:2017ivp}. If the surface is absorbing, however, 
the horizonless object will collapse to a black hole due to accretion of material in the interstellar and in the intergalactic
medium, accretion of dark matter, or absorption of GWs if in a binary system~\cite{Chen:2019hfg,Yunes-prep}. To avoid collapse,
the surface of the horizonless object has to be a sufficiently far from the would-be horizon, possibly
inducing other signatures when two such objects coalesce (see also~\cite{Krishnendu:2017shb} for more discussion). Finally gravitational collapse to horizonless compact objects may involve novel types of GW bursts following the main GW signal \cite{Hertog:2017vod}.
To date, there is no theoretical framework at the level of precise field equations or an action principle usable to study whether these horizonless objects can form from the collapse of matter fields (regardless whether they satisfy the energy conditions). 

\section{Dark Energy and the $\Lambda$CDM Model}
\label{sec:DE-screening}


The observed late-time acceleration of the cosmic expansion is one of the greatest mysteries in modern cosmology. Currently, the tension in the observed value of the Hubble constant from Type IA supernova~\cite{Riess:2016jrr} and the cosmic microwave background~\cite{Ade:2015xua} has only exacerbated this mystery.  
Although by now several observations have undoubtedly confirmed the presence of such anomalous acceleration \cite{Betoule:2014frx,Ade:2015xua,Aghanim:2018eyx}, a convincingly theoretical explanation is still missing \cite{Copeland:2006wr}.
The simplest solution, relying on the addition of the so-called \textit{cosmological constant} into the Einstein field equations, has always been at odds with theoretical expectations \cite{1989RvMP...61....1W,Martin:2012bt} and is now challenged by mounting observational evidence, for example from supernova observations\cite{Riess:2018byc,Riess:2019cxk}.

It is thus not surprising that alternative solutions proliferate in the literature \cite{Copeland:2006wr,Joyce:2016vqv}.
The nature of the so-named \textit{dark energy}, the invisible entity introduced to account for the observed cosmic acceleration, has been the subject of much speculation.
Many different hypotheses have been considered to explain the origin of dark energy, including in particular introducing new cosmological matter fields and modifying GR \cite{Joyce:2016vqv}.
The resulting plethora of dark energy models must be tested against all observations collected so far in order to select the models that are physically viable.
Future observations will further refine this set of viable dark energy models and possibly provide hints towards the solution of the cosmic acceleration riddle.
These future observations include tests with catalogues of GW \textit{standard sirens} collected by LISA.

Standard sirens are GW sources at cosmological distances that can be used as reliable and independent distance indicators, i.e.~which yield a direct measurement of the luminosity distance which does not need to be calibrated with the cosmic distance ladder \cite{Schutz:1986gp,Holz:2005df}.
For cosmological applications they need a corresponding redshift measurement.
Standard sirens can thus be employed either with the joint detection of an EM counterpart, from which the redshift of the GW source can be inferred, or without any EM counterpart identification, in which case the so-called ``statistical method'', which uses galaxy catalogues to infer redshift information, must be applied \cite{Schutz:1986gp,DelPozzo:2011yh}.
LISA will detect mainly three types of GW sources at cosmological distances\footnote{IMBHBs will also be detected at cosmological distances, but their existence is still debated nowadays and consequently no cosmological applications of this class of sources have ever been considered.}: SMBHBs, EMRIs, and SOBHBs sources.
All these sources can be used as standard sirens, although only SMBHBs are expected to provide observable EM counterparts.
Interestingly the subset of these sources that will be relevant for cosmological analyses will be observed at different redshift ranges: SOBHBs will be mainly detected at $z<0.1$ \cite{Kyutoku:2016zxn,DelPozzo:2017kme}, EMRIs at $0.1<z<1$ \cite{MacLeod:2007jd,Babak:2017tow} and SMBHBs at $1<z<10$ \cite{Tamanini:2016zlh}.
This implies that LISA can be considered as a cosmological probe able to test the expansion of the universe 
across the distance ladder~\cite{Tamanini:2016uin}.

All these standard sirens will provide a way to test models of dark energy through the \textit{distance-redshift relation}, which is a standard cosmological relation connecting the luminosity distance to the redshift of any source and it crucially depends on the cosmological model that is chosen, particularly through some model parameters.
A fit of this relation with the standard siren data gathered by LISA will thus provide an effective way to constrain dark energy models, in exactly the same way type-Ia supernova analyses are performed.
Furthermore, many modified gravity models of dark energy, including for example viable Horndeski theories, bigravity and nonlocal gravity (see e.g.~\cite{Maggiore:2013mea,Belgacem:2017cqo}), predict a different distance scaling in the cosmological propagation of GWs.
Specifically the amplitude of GWs does not decrease as simply the inverse of the luminosity distance, as required by GR, but instead it follows a different scaling generally dependent on the redshift through modified gravity terms \cite{Lombriser:2015sxa,Nishizawa:2017nef,Belgacem:2017ihm,Belgacem:2018lbp}.
A similar feature is also predicted by models of dark energy evoking extra spacetime dimensions, where GWs are allowed to propagate in the higher dimensions as well \cite{Deffayet:2007kf,Pardo:2018ipy,Abbott:2018lct}.
In these cases the comparison of the luminosity distance inferred from GW measurements with the luminosity distance as inferred from EM observations, will allow to test deviations from GR and thus to constrain modified gravity models of dark energy.

At present standard sirens are not able to constrain any dark energy model, since currently operating Earth-based detectors can only collect useful cosmological data at low redshift and consequently measure only the Hubble constant \cite{Abbott:2017xzu,Fishbach:2018gjp,Soares-Santos:2019irc,Chen:2017rfc}.
LISA will instead collect cosmological data at higher redshift (possibly up to $z\sim 10$ \cite{Tamanini:2016zlh}), and thus will in fact have the opportunity to test other cosmological parameters, including the ones specifying the nature of dark energy.

Although at the moment there is no complete investigation of the cosmological potential of LISA which takes into account all the possible types of standard sirens mentioned above, some studies considering only SMBHB data with EM counterparts have been performed \cite{Tamanini:2016zlh,Tamanini:2016uin}.
These analyses showed that LISA will effectively constrain dark energy models predicting deviations from the standard $\Lambda$CDM cosmological evolution at redshift $1<z<10$, such as for example early dark energy and interacting dark energy models \cite{Caprini:2016qxs,Cai:2017yww}.
Moreover similar investigations clearly pointed out that LISA will strongly constrain any deviation from the standard GR propagation of GWs, implying that LISA will efficiently test modified gravity models of dark energy \cite{Belgacem:2019pkk}.
To understand if and how LISA will test more conventional dark energy models, where deviations from $\Lambda$CDM appear only at low redshift, further analyses involving the other LISA standard siren sources, namely SOBHBs and EMRIs, are required.
In any case, by probing possible departures from GR and by constraining deviations from $\Lambda$CDM at high redshift, LISA will definitely be able to shed new light on the nature of dark energy. One should also keep in mind that Athena-LISA combined observations might enhance further the possibilities to make progress
on some of the issues mentioned in this white paper~\cite{Athena-LISA-1,Athena-LISA-2}. 

\section{Dark Matter and Primordial Black Holes}
\label{sec:DM-PBH}

LISA observations of rotating black holes could constrain or detect certain light bosonic fields that have been proposed as dark matter candidates~\cite{Marsh:2015xka,Hui:2016ltb}, even in the absence of a direct detection of stochastic GWs of cosmological origin.  The reason is that ultralight bosonic fields around spinning black holes can trigger a superradiant instability~\cite{Detweiler:1980uk,Damour:1976kh,Dolan:2007mj,Shlapentokh-Rothman:2013ysa,Moschidis:2016wew,Frolov:2018ezx,Dolan:2018dqv}
(akin to Press and Teukolsky's ``black hole bomb''~\cite{Press:1972zz}), forming a long-lived bosonic ``cloud'' outside the horizon.  The superradiant instability spins the black hole down, transferring up to a few percent of the black hole's mass and angular momentum to the cloud~\cite{Arvanitaki:2010sy,Brito:2014wla,Arvanitaki:2014wva,Yoshino:2014wwa,Brito:2015oca,Arvanitaki:2016qwi,Baryakhtar:2017ngi,East:2017ovw,Ficarra:2018rfu}.  
The condensate is then dissipated through the emission of GWs with frequency $f\sim m_s/\hbar$, where $m_s$ is the mass of the field.  
Novel GW signatures are expected in binary systems that may be affected by resonant phenomena~\cite{Blas:2016ddr,Baumann:2018vus,Berti:2019wnn,Baumann:2019ztm,Cardoso:2020hca}
or superradiance of the binary system itself~\cite{Wong:2019yoc,Wong:2019kru}.

Superradiance is most effective when the boson's Compton wavelength is comparable to the black hole's gravitational radius~\cite{Dolan:2007mj,Witek:2012tr}.  Strong motivation to investigate this possibility comes e.g. from ``string axiverse'' scenarios (where axion-like particles arise over a broad range of masses in string theory compactifications as Kaluza-Klein zero modes of antisymmetric tensor fields~\cite{Arvanitaki:2009fg}) and from ``fuzzy dark matter'' scenarios (which require axions with masses $\approx 10^{-22}$~eV~\cite{Hui:2016ltb}).
We should note, however, that this effect has even more powerful implications: it is sensitive only to the gravitational interaction with (ultra-light) massive bosons and, thus, it enables us to probe a wide range of beyond-standard model particles in general.

Current Earth-based detectors can probe boson masses $m_s\sim 10^{-13}$--$10^{-11}$~eV, while LISA can detect or rule out bosons of mass $m_s\sim 10^{-19}$--$10^{-15}$~eV~\cite{Brito:2017wnc,Brito:2017zvb,Isi:2018pzk,Ghosh:2018gaw,Tsukada:2018mbp,Baumann:2018vus,Berti:2019wnn}.  
That is, axions in the ``standard'' mass range proposed to solve the strong CP problem of QCD could be tested by GW interferometers on Earth~\cite{Arvanitaki:2014wva,Arvanitaki:2016qwi,Brito:2017wnc}, LISA could test a broad range of masses relevant to string axiverse scenarios, as well as some candidates for fuzzy dark matter. An attractive possibility is the detection of EMRIs around black holes that have formed a boson cloud: the cloud would affect the phase evolution of the system in a characteristic, detectable manner~\cite{Eda:2013gg,Macedo:2013qea,Eda:2014kra,Hannuksela:2018izj,Baumann:2018vus,Berti:2019wnn,Baumann:2019ztm,Naoz:2019pch}, which could be used to identify hypothetical dark matter candidates.

The range of allowed boson masses $m_s$ can also be constrained by LISA measurements of the spins of black holes in binary systems.  For a given $m_s$, black holes should spin down whenever their spin is large enough to trigger superradiant instabilities.  Instability windows in the black hole spin versus mass plane, for selected values of $m_s$, can be obtained by requiring that the instability acts on timescales shorter than known astrophysical processes, such as accretion and mergers.  Roughly speaking, continuum fitting or Iron K$\alpha$ measurements (see e.g. \cite{1112.0172}) of supermassive black hole spins probe the existence of bosons in the mass range $m_s\sim 10^{-19}$--$10^{-17}$~eV. For stellar-mass black holes, the relevant mass range is $m_s\sim 10^{-12}$--$10^{-11}$~eV.  Black hole spin measurements with a space-based GW detector can rule out light dark matter particles in the intermediate mass range $m_s\sim 10^{-16}$--$10^{-13}$~eV, which is inaccessible to electromagnetic observations of stellar and massive black holes \cite{1801.01420}.  Therefore LISA can probe the existence of ultralight bosons in a large mass range that is not probed by other black hole spin measurement methods, or even measure $m_s$ with $\sim 10\%$ accuracy if scalars in the mass range $[10^{-17}, 10^{-13}]$~eV exist in nature~\cite{Brito:2017zvb}.  Spin-one and spin-two fields (i.e., hypothetical dark photons or massive gravitons) would trigger even stronger superradiant instabilities. A space-based detector could detect resolved or stochastic GW from superradiant instabilities, or set strong constraints on the viable mass range for light bosons~\cite{Pani:2012vp,East:2017ovw,Brito:2013wya,Baryakhtar:2017ngi,Barausse:2018vdb,Alexander:2018qzg}.

Self-interacting models of dark matter could also be constrained with LISA observations. Recently, Ref.~\cite{Pollack:2014rja} suggested that (ultra) self-interacting dark matter could form massive seed black holes. These seed black holes would later on grow through accretion to form the supermassive black holes at the centers of galaxies that we see today. Given that LISA can see $10^5 M_{\odot}$ SMBHBs to large distances, it may be possible to probe this scenario through the detection of SMBHB populations at large redshift.

Another interesting candidate for dark matter are primordial black holes (PBHs) \cite{Hawking:1971ei}. In particular, PBHs in the stellar-mass range may contribute a non-negligible fraction of dark matter~\cite{Bird:2016dcv,Clesse:2016vqa,Sasaki:2016jop,Sasaki:2018dmp}.  PBHs can dynamically form binaries, typically resulting in highly eccentric orbits at formation~\cite{Ali-Haimoud:2017rtz}. GW is a direct probe of the self-interaction of PBH dark matter~\cite{Kovetz:2017rvv}.  With its access to earlier stages of the inspiral, LISA could allow us to distinguish the PBH binary formation channel from stellar-origin formation channels through measurements of spin and eccentricity~\cite{Cholis:2016kqi}, as well as the mass spectrum~\cite{Kovetz:2016kpi}. Another source of unique information is through the stochastic background. The PBH merger rate at high redshift is not limited by the star formation rate, and so the stochastic background from these events should extend to lower frequencies (and higher redshifts) than for traditional binary black hole sources~\cite{Mandic:2016lcn,Clesse:2016ajp}.  If PBHs are to form from the collapse of overdense regions deep in the radiation domination era, the required $\mathcal{O}(1)$ fluctuations in the primordial curvature power spectrum will provide a second-order source of primordial GWs~\cite{Ananda:2006af,Baumann:2007zm,Garcia-Bellido:2017aan}. The characteristic frequency of these GWs is directly related to the PBH mass. Interestingly, one of the least constrained mass windows for PBH dark matter (from $10^{-13}\,M_\odot$ to $10^{-11}\,M_\odot$) corresponds precisely to the mHz frequency window accessible by LISA~\cite{Bartolo:2018evs,Bartolo:2018rku,Cai:2018dig}.
It will thus be important to study whether LISA will be able to test the PBH dark matter scenario in this mass window through the two-point and three-point correlations of the GW signal~\cite{Bartolo:2016ami,Bartolo:2018qqn}.

\section{Other Model-Independent Tests}
\label{sec:other-model-indep-tests}

One model independent way of probing GR, which will however require extensive
investigations to see whether it is accurate enough, is to perform a residual test~\cite{TheLIGOScientific:2016src,LIGOScientific:2019fpa}. This is done by subtracting the most probable template from the data and carrying out a Bayesian model selection analysis to see whether the residual is consistent with noise or contains a signal. Such a test can capture not only beyond-GR effects un-modeled in GR template waveforms, but also systematics within GR. LISA is expected to detect GW signals with SNRs much higher than observed events with aLIGO and Virgo. It would be important to repeat the above residual tests to check consistency between LISA data and our GR and waveform expectations.

Another model-independent test is the consistency check among the inspiral-merger-ringdown (IMR) parts of the waveforms~\cite{Ghosh:2016qgn,TheLIGOScientific:2016src,Ghosh:2017gfp,LIGOScientific:2019fpa}, an analysis that resembles a jack-knife test. This approach treats the inspiral and post-inspiral parts of the waveform (separated by the frequency at the innermost stable circular orbit of the binary) independently, and it estimates the remnant's mass and spin with the help of numerical relativity fits. These tests only work for high-mass systems, since one needs to detect \emph{both} inspiral and post-inspiral signals independently. Given that large SNRs are expected for observations of GWs from SMBHBs with LISA, the measurement accuracy of masses and spins of remnant black holes should improve significantly from current measurement of e.g. GW150914~\cite{LIGOScientific:2019fpa}, both with inspiral and post-inspiral parts of the waveform. Therefore, such IMR consistency tests will become more constraining with LISA.

The third model-independent test is to constrain parametric deviations from GR in the waveforms~\cite{Arun:2006yw,Mishra:2010tp,Yunes:2009ke,Cornish:2011ys,Huwyler:2011iq,TheLIGOScientific:2016src,Abbott:2018lct,LIGOScientific:2019fpa}. The parameterized post-Einsteinian (ppE) formalism~\cite{Yunes:2009ke} introduces amplitude and phase corrections to the GR waveform that can capture non-GR effects that do not necessarily follow the GR structure in the waveform, such as corrections entering at negative post-Newtonian (PN) orders\footnote{A correction of $N$ post-Newtonian, or $N$PN, order is one that is of ${\cal{O}}(v^{2N}/c^{2N})$ relative to the GR leading order term~\cite{Blanchet:2013haa}.}. The LVC used the generalized IMRPhenom (gIMR) model~\cite{TheLIGOScientific:2016src}, which also includes corrections in the merger-ringdown part of the waveform\footnote{The original ppE formalism also proposes a similar way to parameterize non-GR modifications in the ringdown phase~\cite{Yunes:2009ke}}. In the inspiral part, the ppE parameters have a one-to-one correspondence with the gIMR parameters~\cite{Yunes:2016jcc}. Bounds on these generic non-GR parameters can be mapped to test the fundamental pillars of GR, such as SEP and LLI~\cite{Yunes:2016jcc}. Reference~\cite{Chamberlain:2017fjl} derived projected bounds on the ppE parameters for a variety of LISA sources and found that EMRIs would provide the most stringent constraints. Another interesting possibility is to perform a parameterized test of GR by combining LISA observations of a stellar-mass black hole binary with ground-based ones of the same source~\cite{Sesana:2016ljz,Sesana:2017vsj,Tso:2018pdv,Gerosa:2019dbe,Cutler:2019krq,Moore:2019pke}. Such multiband GW observations allow us to constrain the ppE parameters more stringently than ground-based detectors alone~\cite{Barausse:2016eii,Vitale:2016rfr,Carson:2019rda,Gnocchi:2019jzp,Carson:2019kkh,Carson:2019yxq}.

A fourth model-independent test is to probe non-GR polarization modes of GWs~\cite{Chatziioannou:2012rf,Abbott:2017oio,Abbott:2018lct,LIGOScientific:2019fpa}. GR only contains two tensorial polarizations, while metric theories of gravity in general can have additional scalar and vector polarization modes (two each)~\cite{Will:1993ns,lrr-2006-3}. Reference~\cite{Tinto:2010hz} derived sensitivities to such additional polarization modes with LISA. They found that sensitivities to vector and longitudinal (transversal) scalar modes are higher than (comparable to) those for the tensor modes. On the other hand, sensitivities of LISA to circular-polarization modes (due to e.g. parity violation in gravity) in stochastic GW background were discussed in~\cite{Seto:2006hf,Seto:2006dz,Smith:2016jqs}.

Other model-independent tests include probing the propagation speed of GWs (comparing to electromagnetic counterparts) or the existence of scalar dipole radiation. The latter is a 
common feature of non-GR theories with additional degrees of freedom, including scalar-tensor theories~\cite{Berti:2004bd}, Einstein-dilaton-Gauss-Bonnet (EdGB) gravity~\cite{Yagi:2011xp} and vector-tensor theories~\cite{Foster:2006az,Foster:2007gr,Blas:2011zd,Yagi:2013ava,Hansen:2014ewa}. One can model such a correction by adding a formally $-1$PN term to the GW energy flux in GR. Observations of GW150914-like stellar-mass binary black hole GWs with LISA will be able to place stringent bounds on dipole emission~\cite{Barausse:2016eii}. A possibility of using GW observations of galactic binary WDs with LISA is discussed in~\cite{Littenberg:2018xxx}, since the presence of dipole emission would effectively make these GWs not monochromatic.   

\section{Astrophysical Systematics}
\label{sec:astro-sys-and-stacking}

Unlike sources for LIGO/Virgo, many of the black hole systems targeted by LISA are believed to live in matter-rich environments~\cite{Barausse:2014pra,Barausse:2014tra}. 
The reason is that massive black holes are supposed to spend at least roughly $1-10$\% of their cosmological evolution in an 
AGN phase~\cite{Shankar:2011mc,2016ApJ...831..203P,2016ApJ...826...12J}, where
they are expected to be surrounded by accretion disks. 

For binaries involving two massive black holes, these accretion disks, even if they were to survive unscathed to the last stages of the binary evolution, are
unlikely to affect the orbital evolution of the system in a detectable way, and hence they should not introduce significant systematic errors in
the GW measurements. However, the circumbinary part of the accretion disk may anchor relatively strong magnetic fields, which coupled with
the motion of the black hole binary and with black hole spins may trigger the launch of powerful electromagnetic jets~\cite{Palenzuela:2010nf,Giacomazzo:2012iv}. The latter may be observable by future 
radio telescopes that will be operational at the same time as LISA (e.g. SKA) and may thus provide a way to better localize the binary in the sky and possibly measure its redshift (either directly via the 21 cm line, or by identifying the host galaxy)~\cite{Tamanini:2016zlh}.

For EMRIs, the accretion disk potentially surrounding the central massive black hole is unlikely to be destroyed by the satellite, although the latter may carve a gap
in the disk (depending on the disk's parameters -- especially its height and viscosity -- and the satellite's). Depending on the system's parameters, 
the satellite may undergo either type I or type II planetary migration, which may result in a dephasing of several radians during the LISA observation span~\cite{Yunes:2011ws,Kocsis:2011dr,Kocsis:2012cs,Kocsis:2012ui,Barausse:2014pra,Barausse:2014tra,2018arXiv181003623D}. Therefore, this effect is likely to be measurable (at least in the $1-10$\% of EMRIs that are expected to involve an AGN) and if unaccounted for may even bias
the measurement of the other parameters. Other potentially measurable effects, though with magnitude slightly lower than that of planetary migration, include dynamical friction 
and hydrodynamic drag from the disk (especially for inclined and/or counter-rotating orbits), and accretion onto the satellite and the central black hole~\cite{Barausse:2007dy,Barausse:2014pra,Barausse:2014tra}.

EMRIs form deep in the cluster that surrounds the massive black hole in galaxy centers.  In such clusters, there will always be other stellar-mass bodies near the EMRI.  These bodies will tidally perturb the EMRI spacetime, and thus perturb the smaller body's orbit.  For most of an EMRI's inspiral, the impact of nearby perturbers is negligible, since the effect of the tidal perturbation averages away over an orbit.  However, in every inspiral there will be multiple moments at which the inspiral's orbit resonates with the tidal distortion of the perturber.  During such tidal resonances, the tidal perturbation does not average away, but instead kicks the system until it evolves out of resonance.  Recent estimates \cite{Bonga:2019ycj} show that this will have a several radian impact on the waveform.  More work is needed to examine how this will affect EMRI measurements, but it is certain to bias GW parameter estimation.  Since tidal resonances cannot be predicted in advance but depend on the (random) distribution of stellar-mass objects near each EMRI, this will be a very difficult systematic to model.

Finally, especially if EMRIs exist in dwarf, dark-matter dominated galaxies, the dark-matter profile around the central black hole may develop  a steep cusp~\cite{Gondolo:1999ef},
which may survive to low redshifts (where EMRIs are detectable) since dwarf galaxies only experience rare mergers (which may destroy the cusp), have low stellar to dark matter mass ratios (which is relevant since stars can destroy the cusps) and short relaxation times (i.e. short cusp regrowth times). Therefore, if these cusps exist, they may leave an observable imprint on these EMRIs (via direct gravitational pull, dynamical friction, hydrodynamic drag and accretion), and possibly bias the GW parameter estimation if not included in the waveform model~\cite{Barausse:2014pra,Barausse:2014tra,Eda:2014kra}. 

In more conventional 
dark matter scenarios, however, the effect of dark matter is not expected to be observable with LISA and/or cause significant systematics~\cite{Barausse:2014pra,Barausse:2014tra}. 
One notable exception, however, is provided by dark matter made of axions or boson ~\cite{Hui:2016ltb}, i.e. the possibility that the dark matter may consist of an ultralight axion-like scalar field. As we discussed in Sec.~\ref{sec:DM-PBH}, in such a scenario, the scalar field may form rotating dipolar condensates around spinning massive black holes, as a result of superradiance. These
rotating dipoles would produce monochromatic GWs~\cite{Okawa:2014nda}
that would be detectable by LISA as resolved sources or as a stochastic background~\cite{Brito:2017zvb,Brito:2017wnc}. Moreover, the gravitational pull of these condensates on EMRIs would affect the phase evolution of the system in a characteristic, detectable manner~\cite{Eda:2013gg,Macedo:2013qea,Eda:2014kra,Baumann:2018vus,Hannuksela:2018izj,Berti:2019wnn,Naoz:2019pch}, 
as we discussed in Sec.~\ref{sec:DM-PBH}.

Very little to no work has gone into the study of how astrophysical systematics could affect our ability to test GR. One could imagine that if astrophysical effects are not accounted for, one may confuse a GR deviation with an astrophysical deviation. If so, statistical methods could come to our 
rescue. If GR deviations are truly present in the signals, one would expect they would be present in all events, and not just in a subset that is prone to be contaminated by astrophysical systematics. If astrophysical effects can be modeled, one could further study whether they can be disentangled from GR deviations. This would be possible if the GR deviations modify the waveform at high PN order predominantly, while astrophysical effects enter first at lower PN order (as is the case, e.g. when considering migration of type I or II in EMRIs with an accretion disk).   


\section{Waveform Systematics}
\label{sec:waveform-sys}

To use GWs as a tool to test the nature of gravity and the fundamental laws of physics, we need models which faithfully represent the predictions of these laws.  The precision with which a theory's predictions can be measured is ultimately limited by how well one can model these predictions.

Phase counting arguments teach us that models used as templates for measuring a particular source must meet certain phase accuracy requirements.  To be useful as {\it detection templates}, a model must match the phase of nature's signal to within a phase error $\delta\Phi \sim 1$ radian; to be useful as a {\it measurement template}, it must match to within $\delta\Phi \sim 1/{\rm SNR}$.  ``Detection templates'' are waveform models that we use to demonstrate that a GW signal is present in detector noise.  It is not necessary that a model faithfully represents the astrophysical signal --- it may have significant systematic errors, but still match phase well enough that a signal can be found in noise.  ``Measurement templates'' are waveform models whose phase is computed accurately enough that we can be confident that intrinsic systematic errors are smaller than the statistical uncertainty associated with measurement noise.  Rigorous calculations backing up these rules of thumb can be found in Refs.\ \cite{Vallisneri:2007ev,Lindblom:2008cm,Vallisneri:2011ts}.

When thinking about using GWs as a tool for probing the fundamental laws of physics, one is ultimately limited by systematic modeling effects. For example, the simultaneous presence of multiple sources in the LISA data stream can make it difficult to extract fundamental physics from weak signals that are overshadowed by much stronger ones. That is, we expect to measure GWs emitted by (super-) massive BH binaries with an SNR of $\mathcal{O}(10^2\ldots10^3)$, so current modeling accuracy may suffice for identification and 
reasonable ($10\%$) error in parameter estimation. However, these modeling requirements will also crucially affect our ability to extract and interpret signals contained in the residuals after the subtraction of such loud events, such as signals from EMRIs or SOBHs, or hints of new physics.
Furthermore, the binaries' expected configurations will probably differ from that of LIGO sources.
In particular, LISA will listen to eccentric binaries 
(that, in triple-systems can reach $\gtrsim0.9$~\cite{Bonetti:2018tpf}),
intermediate mass ratios, and highly spinning BHs.

The different sources discussed in this paper require diverse waveform modeling techniques, each with their own challenges.
In practice, this Herculian task offers great opportunities for synergies and collaboration
between different Working Groups of the LISA Consortium.
For example, the accuracy requirements needed to search for deviations from our standard models of gravity, particle physics and 
cosmology will drive the development of innovative modeling techniques in GR and beyond. For the purpose of this paper, let us specify three sets of techniques
to extract the most information about fundamental physics from LISA data: 
\begin{itemize}
\item {\textbf{Post-Newtonian (PN) methods:}}
These analytic techniques are a series expansion in weak fields and small velocities, yielding solutions that describe the dynamics of a binary well during the inspiral.
PN is an established tool that has been used to produce two of the main waveform models deployed in LIGO data analysis.
The accuracy of the description is controlled by the PN order, which identifies the relative order to which an expansion has been taken in small velocities and weak fields\footnote{A term that scales as $(v/c)^{2N}$ is said to be of N PN order relative to its leading-order controlling factor.}. While waveforms of non-spinning, circular binaries are currently known up to 4PN order, those of eccentric, spinning or precessing binaries are currently only available up to 3PN and 3.5PN order. 

Possible advances that will lead to extensions to higher PN order are under way. These include novel connections to EMRI modeling~\cite{Bini:2015bfb}, the use of effective field theory and particle physics techniques~\cite{Goldberger:2004jt}, and the further development of Hamiltonian and Lagrangian methods coupled to dimensional and Hadamard regularization~\cite{Marchand:2017pir,Damour:2016abl}. 

\item {\textbf{Numerical relativity (NR):}}
These numerical techniques solve the field equations through high-performance computing, yielding solutions that describe the dynamics of a binary during the merger.
NR is an established tool that has produced extensive waveform catalogues (for LIGO)~\cite{Boyle:2019kee,SXSCatalogue:web,GATechCatalogue:web,RITCatalogue:web}.
Yet, we face new challenges for LISA's source modeling that go far beyond  ``simply'' increasing resolution to decrease numerical discretization errors.
The latter demands major code enhancements to optimize performance and increase scalability to exa-scale high-performance computing facilities,
that has to go hand-in-hand with novel theoretical techniques.

Possible advances include new, constraint preserving formulations and appropriate gauge choices that ensure long-term stable and accurate solutions.
Different avenues to address these challenges are underway.

\item {\textbf{Waveform modeling:}}
  The construction of a final waveform model that covers the signal from the early inspiral all the way through the merger and ringdown will probably require the construction of phenomenological~\cite{Husa:2015iqa,Khan:2015jqa} and of  effective-one-body (EOB) techniques~\cite{Buonanno:1998gg,Buonanno:2000ef}. The former ``stitches'' PN and ringdown waveforms together through the inclusion of higher PN order terms that are fitted to numerical simulations. The latter ``stitches'' a resummed version of the PN expansion to a ringdown waveform through the inclusion of higher PN order terms in the Hamiltonian and radiation-reaction force that are fitted to numerical simulations. These complementary methods are conceptually well established, but, their accuracy is ultimately
limited by the present knowledge of their PN and numerical relativity components.
\end{itemize}
This list refers to upcoming tasks within GR. Any extension thereof -- be it in a theory-specific or theory-agnostic fashion -- requires major re-thinking of established concepts that is still in its infancy.

GWs from EMRIs are interesting for thinking about systematic effects, since they are a source class for which modeling is largely understood in principle, but for which practical considerations make modeling challenging in practice.  Consider the following modeling challenges:
\begin{itemize}

\item {\bf Self force order.}  EMRIs are modeled by perturbing exact black hole spacetimes, and using the perturbation to calculate a ``self force'' \cite{Poisson:2011nh,Barack:2018yvs} that pushes the small body's motion away from leading black hole orbit.  The perturbation is organized in powers of the EMRI mass ratio $\mu/M$.  A phase counting argument \cite{Hughes:2016xwf} suggests that measurement templates will require us to go to at least second order in this expansion, a research program that is just getting underway.

\item {\bf Structure of the smaller body.}  Most EMRI waveform models treat the smaller body as a point-like mass, ignoring the fact that it has finite extent and some form of internal structure.  This body of course has some structure, which will couple to the background spacetime; see \cite{Witzany:2019dii} for recent discussion and review.  This coupling causes the small body's spin to precess, and exerts a force (relative to the motion of a non-spinning body) which changes the small body's motion.

\item {\bf Resonances.}  Most of the small body's inspiral phase arises from the orbit-averaged, dissipative effect of the self force, which is equivalent to the loss of energy and angular momentum due to GW emission.  In every inspiral, there will be moments at which two of the orbit's orbital frequencies are commensurate.  At these moments, terms which normally average to zero will fail to average away.  At such resonant moments, the EMRI's evolution can change quite a bit \cite{Flanagan:2010cd, Flanagan:2012kg}.  This is very similar to the tidal resonance discussed in Sec.\ \ref{sec:astro-sys-and-stacking}.  How an EMRI's evolution changes in detail depends on the relative phase of the orbit's radial and angular motions as it enters the resonance, much as the tidal resonance depends on the random distribution of perturbing stellar-mass objects near the EMRI.

\end{itemize}

It is worth emphasizing that these effects refer exclusively to EMRIs and they are difficult to model, but they are present in GR.  It will be challenging to incorporate additional subtle effects that arise in different theories of gravity, and to understand how such effects may be disentangled from all other properties of the source with which they may be correlated.  Resonance effects are particularly worrisome from the standpoint of waveform systematics.  Because they depend on system characteristics that cannot be predicted in advance (the phase at which the system enters resonance; the distribution of nearby bodies), they may provide the ultimate systematic limitations on waveform phase precision. Precisely how such limitations percolate into limits on how well we can probe fundamental physics remains an open problem.

The above description should not be construed as implying that the only GW sources that may contain waveform systematics are EMRIs. In fact, similar modeling problems arise when considering other sources of GWs, such as SMBHBs and IMBHBs. The latter, for example, are sources in which the PN approximation is not as accurate as when mass ratios are comparable (see e.g.~\cite{Yunes:2008tw}). Perturbation theory techniques are also not very accurate for such sources because the mass-ratio is not as small as when considering EMRIs. Full numerical simulations are currently computationally challenging because large mass ratio binaries take longer to inspiral and require a larger dynamical range in adaptive mesh refinement. Without PN, perturbation theory or numerical waveforms, it becomes extremely challenging to validate resummed PN waveforms, such as those coming from the effective-one-body framework~\cite{Buonanno:1998gg,Buonanno:2000ef}.  

\section{Conclusions}
\label{sec:conclusions}

LISA has immense potential to test for GR deviations in the extreme gravity regime, where the gravitational interaction is enormous and curvatures are large and dynamically changing during the observation time. Such probes of fundamental physics will inform theoretical studies to resolve outstanding questions, of both theoretical and observational nature, in our standard gravitational and cosmological models. Although a lot of the work done so far has dealt with perturbative modifications to the predictions of GR, we note that also global non-perturbative modifications of the geometry may affect both the generation and propagation of GWs, which in turn could constrain quantum gravity models (see e.g.~\cite{Calcagni:2019kzo,Calcagni:2019ngc}). 

We have here laid out the main physics drivers for such fundamental physics probes together with their current state of development. The goal of this paper is thus to serve as a guide and a reference for the community of people that may be interested in pursuing fundamental physics problems with LISA. Given that the physics drivers may change and evolve with time, the content of this paper should in some sense be understood as live and changing. We trust that this will constitute a basis for further studies of LISA science related to theoretical physics thereby strengthening the LISA scientific community as a whole. 

\acknowledgments We would like to thank all of the attendees of the first, inaugural meeting of the Fundamental Physics Research Group in Florence for interesting and stimulating discussions and presentations. We would also like to thank the Galileo Galilei Institute for their hospitality during the organization of this meeting.
E. Barausse, A. Dima, N. Franchini and S. V\"olkel  acknowledge financial support provided under the European Union's H2020 ERC Consolidator Grant
``GRavity from Astrophysical to Microscopic Scales'' grant agreement
no. GRAMS-815673.
This work has also been supported by the European Union's Horizon 2020 research and innovation program under the
Marie Sklodowska-Curie grant agreement No 690904.
The authors would like to acknowledge financial and networking support by the
GWverse COST Action CA16104, ``Black holes, gravitational waves and fundamental
physics''; the European Research Council Starting Grant DarkGRA-757480 (``Unveiling
the dark universe with gravitational waves'') and the Flemish Research Council through the Odysseus grant  G.0011.12.
H. Witek acknowledges financial support provided by the Royal Society
University Research Fellowship UF160547 and the Royal Society Research
Grant RGF\textbackslash R1\textbackslash 180073.
S.A. Hughes's work on LISA-related science is supported by NASA Grant
No.~80NSSC18K1091.
E. Berti is supported by NSF Grants No. PHY-1912550 and AST-1841358, NASA ATP Grants
No. 17-ATP17-0225 and 19-ATP19-0051, and NSF-XSEDE Grant No. PHY-090003.
N.Yunes acknowledges support from NSF grant PHY-1759615, and NASA grants
80NSSC18K1352.
K. Yagi acknowledges support from NSF Award PHY-1806776, a Sloan Foundation
Research Fellowship, and the Ed Owens Fund.
T. P. Sotiriou acknowledges partial support from the STFC
Consolidated Grant No. ST/P000703/1.


\bibliography{refs}

\end{document}